\documentstyle[prl,aps,twocolumn]{revtex}
\large
\begin{document}
\draft
\twocolumn[\hsize\textwidth\columnwidth\hsize\csname @twocolumnfalse\endcsname

\title{Kondo effect versus Jahn-Teller effect: A new class of quantum
impurity problems}
\author{
Alexander O. Gogolin
}
\address{
Institut Laue-Langevin, B.P.156 38042 Grenoble, Cedex 9, France\\
and\\
Landau Institute for Theoretical Physics, Kosygina str. 2,
Moscow, 117940 Russia
}
\date{\today}
\maketitle
\begin{abstract}
Impurity ions with orbitally degenerate ground electronic configuration
embedded in metals are subjects to both the Kondo effect and the Jahn-Teller
instability. While increasing the Jahn-Teller coupling, one passes,
from the region of parameters where the perturbation theory is
applicable, through a complicated intermediate region, to
a region where phonons, localized electrons and conduction
electrons form a coherent state.
\end{abstract}
\pacs{ }
]
\narrowtext

In the course of time, the Kondo model, originally proposed in
an attempt to account for a resistance minimum in magnetic
alloys, became a kind of reference model in
condensed matter theory,
revealing nontrivial infrared properties of metals.
It is the simplest, single channel, version of the Kondo model
that has required decades for its theoretical understanding
serving as a laboratory for applying
different methods\cite{single-Kondo}.
Along with this development, though somewhat delaying,
efforts to devise realistic models describing magnetic
impurities in metals were taken.
These investigations date
from the discussion of an impurity ion $d$-level degeneracy
in the original Anderson paper\cite{Anderson} and
(after \onlinecite{Goqblin-Schrieffer} and a group theoretical
analysis of \onlinecite{Hirst}) they led to
the paper \onlinecite{Boss2}
where a class of models, now known as multi channel Kondo
models, has been sorted out and an attention to
a peculiar (``non Fermi liquid") low energy behaviour
of these models was drawn.
Bethe ansatz solutions for multi channel Kondo
models were shortly reported\cite{Bethe-multi}.
During last few years,
numerous studies of the unconventional low
energy features have been carried out.
They include large $k$ ($k$ - number of channels)
expansion\cite{large-k}, conformal field theory\cite{CFT} and
bosonization\cite{bosonization} approaches.
Despite this essential success in studying the models,
their relevance to concrete impurity problems has not been
explored in detail, basically remaining on the level of
Ref.\onlinecite{Boss2}.
A remarkable exception to this is given by comprehensive investigations
of two-level systems in metals whose low energy behavior was
conjectured to be described
by the two channel Kondo model\cite{Zawadowski}; the
conjecture being recently confirmed by
point contact spectroscopy experiments\cite{PCS}.

One of the main sources of motivations for studying multi-channel
Kondo models is provided
by unusual temperature (magnetic field) dependence of various
physical quantities in heavy fermion compounds\cite{hf-general}.
This issue has been raised by Cox for the case of Uranium
compounds\cite{Cox1} and, recently, for some Cerium
compounds\cite{Cox2}.
At least for dilute systems, when coherence effects can be neglected,
single impurity properties are likely to be dominants
(though, even for dense compounds,
there are considerable empirical evidences favouring local physics).
Fortunately, new (and
puzzling) experimental data on dilute $U$ impurities in
$MPd_3$ (with $M$ being $Y$, $Sc$, $La$, etc.) became recently
available\cite{Maple}. Fermi level tunning\cite{JimAllen}
and ionic radii\cite{Maple} arguments evidence $U$ being in $U^{4+}$
ionic state.
The $J=4$ Hund's rule ground state is split by the cubic
crystal field into $\Gamma_1$, $\Gamma_3$, $\Gamma_4$, and $\Gamma_5$
states.
Admitting that the point charge model of
Ref.\onlinecite{splittings} is oversimplified,
one may assume\cite{Cox1,Maple} the $\Gamma_3$ nonmagnetic
doublet being the ground state
(in agreement with neutron scattering data\cite{neutron}
though the latter experiments are performed on rather heavily
doped samples).
That would lead to a two channel Kondo model explaining
part of the data\cite{Maple} - the logarithmic low temperature
specific heat and the magnetic susceptibility
behaviour\cite{Cox3} but not
the linearly (in temperature) saturating resistivity which
remains to be reconciled.
This line of arguments concerns purely electronic
properties disregarding lattice degrees of freedom.
An important point is, however,
that an orbitally degenerate ground state of
the impurity ion realizes classical conditions for the
Jahn-Teller instability\cite{Abraham}.
Crucial difference with the standard situation is that local
phonon modes interact here not only with localized electrons but
also with those from conduction band (coupled, in turn, to the
degenerate local electronic configuration).
Apart from some brief comments (e.g., in Ref.\onlinecite{Cox4}),
this interesting problem has been overlooked in the
literature (and poorly understood).

This letter is intended to give a qualitative physical discussion
of that complicated interplay of the Kondo- and the Jahn-Teller
effects realized by orbitally degenerate ions in metals where
the three ingredients of the system (the local electronic configuration,
the conduction electrons, and the local phonon modes) are
tied together. The main trends are explored whereas
detailed calculations (anyway not allowed by the format of the letter)
will be presented elsewhere\cite{elsewhere}.

The problem exists for any orbitally degenerate ionic state
($\Gamma_3$, $\Gamma_4$, $\Gamma_5$ and $\Gamma_8$ for the cubic
point group), magnetic or not.
Low energy properties are of interest, so the temperature
is understood to be small as compared to the crystal field
splitting energy (on the scale of $100K$ for $U-MPd_3$ if
estimated from the specific heat Schottky anomaly\cite{Maple}).
Generally, the Hamiltonian of the problem can be written in the form:
\begin{equation}
H=H_{el}+H_{ph}+H_{el-ph}+H_{K}+H_{JT},
\label{GeneralHamiltonian}
\end{equation}
where $H_{el}$ describes conduction electrons,
$H_{ph}$ accounts for the lattice,
$H_{el-ph}$ represents conventional electron-phonon coupling,
and $H_{K}$ and $H_{JT}$ stand respectively for the Kondo- and
the Jahn-Teller couplings to be specified shortly.

As an example the cubic
point group and the $\Gamma_3$ state is considered in this letter.
The $\Gamma_3$ representation is spanned by localized electronic
wave functions transforming as $|\theta\rangle\propto 3z^2-r^2$
and $|\varepsilon\rangle\propto x^2-y^2$.
$[\Gamma_3^2]=\Gamma_1+\Gamma_3$ and the operators (again spanning
$\Gamma_3$), which describe mixing of $|\theta\rangle$ and
$|\varepsilon\rangle$ due to an interaction with the environment,
are: $U_\theta=|\varepsilon\rangle\langle\varepsilon|-
|\theta\rangle\langle\theta|$
and $U_\varepsilon=|\theta\rangle\langle\varepsilon|+
|\varepsilon\rangle\langle\theta|$.
[Additionally there is $U^2=i(|\theta\rangle\langle\varepsilon|-
|\varepsilon\rangle\langle\theta|)$ operator belonging to
$\{\Gamma_3^2\}=\Gamma_2$ representation.]
These are
essentially Pauli matrices, so the $\Gamma_3$ doublet may
be thought of as a pseudospin (not transforming, of course,
as a true spin under spatial rotations).

For the Kondo coupling, one may therefore write:
\begin{equation}
H_{K}=U_\theta S_\theta +U_\varepsilon S_\varepsilon+U^2 S^2,
\label{GeneralKondocoupling}
\end{equation}
where $S_\theta$ and $S_\varepsilon$ are such combinations of the
operators describing conduction electrons scattering from the
impurity ion which span the $\Gamma_3$ representation [for
(\ref{GeneralKondocoupling}) to be invariant].
Due to symmetry arguments,
those scattering operators, spanning in general some representation
$\Gamma_\alpha$, can be constructed as:
\begin{equation}
S^{\alpha}_p=\sum_{{\bf \hat{k}},{\bf \hat{k'}}}
\sum_{\beta q, \gamma r}
J^{\alpha}_{\beta\gamma}({\bf \hat{k}},{\bf \hat{k'}})
\langle \alpha p || \beta q, \gamma r \rangle
c^{\dagger}_{{\bf \hat{k}},\beta q}
c^{\phantom{\dagger}}_{{\bf \hat{k'}},\gamma r}.
\label{Generalscatteringoperators}
\end{equation}
Here the operators
$c^{\phantom{\dagger}}_{{\bf \hat{k}},\gamma r}$ are defined
in a standard way as appropriate linear combinations of
operators
$c_{\bf k}$ belonging to the star of the momentum vector ${\bf k}$;
${\bf \hat{k}}$ numbers all the stars,
indexes $\alpha$, $\beta$ and $\gamma$ denote
irreducible representations ($p$, $q$ and $r$ standing for
degenerate states spanning these representations),
$\langle \alpha p || \beta q, \gamma r \rangle$ are corresponding
Clebsh-Gordan coefficients, and
$J^{\alpha}_{\beta\gamma}({\bf \hat{k}},{\bf \hat{k'}})$
are the coupling constants.
It should be sad that the expression (\ref{Generalscatteringoperators})
contains, strictly speaking, infinite number of orbital channels.
Yet only those few channels which have the largest coupling
constants survive the scaling, the other ones tend to die
out in the course of the renormalization
group transformations\cite{Boss2,bosonization,Zawadowski}.
It is this tendency which makes the problem tractable.
The determination of relevant channels is a complicated problem
which requires a detailed analysis of the concrete impurity structure
and which presently is far from being clear.
For the case of the (nonmagnetic) $\Gamma_3$ doublet,
$S_{\theta,\varepsilon}$ is a sum of time-reversal partners,
so that the number of orbital channels has to be doubled.
Hence, in the simplest case, one arrives at the two channel model,
as proposed in \cite{Cox1}.

The Jahn-Teller coupling is represented by:
\begin{equation}
H_{JT}=g_{JT}\left( U_\theta Q_\theta
+U_\varepsilon Q_\varepsilon \right),
\label{Jahn-Tellercoupling}
\end{equation}
where $\left( Q_\theta ,Q_\varepsilon\right)$ is a local phonon mode
transforming as $\Gamma_3$. The relevant phonons are understood
to be optical ones and their dispersion is neglected (it is not
expected to qualitatively modify the results).
In general there is also a non-linear coupling (along with
phonons anharmonicity).

In order to proceed, it is instructive to formally ``integrate out"
electronic degrees of freedom. This can be done in a standard way
with the following result for the effective imaginary time
action for the phonon mode:
\begin{eqnarray}
&~&S\left\{Q\right\}=S^{(0)}\left\{Q\right\}-
\label{effectiveaction}\\
&~&\int_{0}^{g_{JT}}dg\int_{0}^{\beta}d\tau
\left[ M^{(g)}_\theta (\tau)Q_\theta(\tau)
+M^{(g)}_\varepsilon(\tau)Q_\varepsilon(\tau)\right],
\nonumber
\end{eqnarray}
where $S^{(0)}\left\{Q\right\}$ is the free action,
$\beta$ is the inverse temperature, and
\[
M^{(g)}_{\theta,\varepsilon}(\tau)=-\langle U_{\theta,\varepsilon}(\tau)
\rangle_{g_{JT}=g}
\]
is the exact non-linear dynamic ``magnetization" of the $\Gamma_3$
pseudospin.
Eq.(\ref{effectiveaction}) contains both the
electronic adiabatic potential for $Q_{\theta,\varepsilon}$ and a
retardation term originating from the long time relaxation
of low energy electronic excitations.

Consider the adiabatic potential first. It arises from the static
magnetization (i.e. calculated for $\tau$-independent
$Q_{\theta,\varepsilon}$).
As it is evident from the structure of (\ref{GeneralKondocoupling})
and (\ref{Jahn-Tellercoupling}), $M^{(g)}_{\theta,\varepsilon}
=(Q_{\theta,\varepsilon}/Q)M(gQ)$ with $Q^2=
Q_{\theta}^2+Q_\varepsilon^2$, $M$ being the transverse
magnetization.
Though an explicit analytic expression for $M$ is generally
lacking, in limiting cases it can easily be determined.
{}From essentially free pseudospin magnetization in high fields
(temperatures) it changes to a well known logarithmic behaviour
in low fields (temperatures). Concerning the cross over,
a set of numerical data for a spin isotropic case\cite{Sacramento}
and analytic formulae at the Emery-Kivelson line\cite{bosonization}
are available.
The electronic part of the adiabatic potential adds a negative
contribution to the elastic energy $(1/2)\mu\omega_0^2 Q^2$,
where $\mu$ is the mass and $\omega_0$- the bare frequency
for the $Q$-mode.
Hence the equilibrium position $Q=0$ may become unstable and
shift to a finite $Q=Q_0$.
By differentiating the adiabatic potential in $Q$ one arrives
at the following equation for $Q_0$:
\begin{equation}
\mu \omega_0^2 Q_0 = g_{JT}M(g_{JT}Q_0).
\label{Qzero}
\end{equation}
It is worth noticing that, since the magnetization should be a
monotonic function of the field, this equation can have only one
nontrivial ($Q_0\neq 0$) solution that appears at temperatures
below $T_{JT}$ which is determined by the following condition:
\begin{equation}
\mu \omega_0^2 =g_{JT}^2 \chi (T_{JT}),
\label{JahnTellertemperature}
\end{equation}
where $\chi$ is the linear susceptibility.
If the Kondo coupling (\ref{GeneralKondocoupling}) is disregarded
[as well as the electron-phonon coupling in
(\ref{GeneralHamiltonian})] then a conventional Jahn-Teller effect
takes place and $T_{JT}$ is related to the energy gain due to the
zero-temperature lattice distortion $W_{JT}=g_{JT}^2/2\mu\omega_0^2$
(actually $T_{JT}=2W_{JT}$).
In the case when this energy is larger
then the characteristic Kondo energy
scale, $T_K$, the Kondo coupling (even if kept finite) does not
essentially influence the formation of the lattice distortion.

One therefore arrives at the following conclusion:
the relative importance of the Kondo- and the Jahn-Teller couplings
is determined by the interplay of the parameters $W_{JT}$
and $T_K$.

Assume that
\begin{equation}
W_{JT}\ll T_K
\label{relative}
\end{equation}
(i.e. the Jahn-Teller coupling is weak). Then one can
expect $T_{JT}\ll T_K$ so that the low temperature
susceptibility $\chi(T)=(1/2\pi T_K)\ln (T_K/T)$ may be used to
estimate
\begin{equation}
T_{JT}=T_K\exp\left\{ -2\pi\mu \omega_0^2 T_K/g_{JT}^2\right\}.
\label{JTtemperature}
\end{equation}
At this temperature a nonzero $Q_0$ appears first, then it increases
with lowering the temperature reaching $T=0$ value which, since a
magnetic field scales as $\sqrt{T}$, is estimated as
\begin{equation}
Q_0= \frac{T_K}{g_{JT}}\sqrt{\frac{T_{JT}}{T_K}}.
\label{distortion}
\end{equation}
The energy gain $\tilde{W}_{JT}$
due to this lattice distortion differs
from the bare $W_{JT}$; it is determined by the energy scale
(\ref{JTtemperature}): $\tilde{W}_{JT}=T_{JT}/4\pi$.
It should be emphasized that for the Jahn-Teller effect
to be fairly developed, the lattice distortion (\ref{distortion})
should exceed the characteristic scale for
zero-point fluctuations around the new minimum:
\begin{equation}
\sqrt{\langle (\Delta Q)^2\rangle} \ll Q_0;
\label{zero-point}
\end{equation}
otherwise the Jahn-Teller coupling does not lead to any significant
effects (does not really cause a distortion) and can be treated
by perturbation theory\cite{noteperturbation}.

Theoretically, the case when both inequalities (\ref{relative})
and (\ref{zero-point}) are fulfilled is probably the most interesting
one - because of its dynamics.
The system can tunnel between different
equivalent minima in the $Q=Q_0$ manifold. Beyond the adiabatic
approximation, each tunneling event causes long relaxation process
involving an adjustment of the electronic state (the ``Kondo
cloud") to a new phonons configuration. This is somewhat
similar to what was proposed by Yu and Anderson\cite{YuAnderson}
for $A15$ compounds, though here the situation is much more complicated
since a dynamic response of the two channel Kondo system to
a rotating (or jumping by $2\pi/3$ depending on the importance of
warping terms) magnetic field comes into play\cite{elsewhere}.
One should mention
that, since the lattice distortion arises from a competition
of two quadratic in $Q$ energies (elastic and electronic parts of the
adiabatic potential) the resulting minimum is going to be
rather flat so that
the inequality (\ref{zero-point}) leads to an exponential constraint
on the mass of the $Q$-mode:
\begin{equation}
\mu \gg g_{JT}^2/T_K T_{JT}.
\label{constraint}
\end{equation}

Turning now to the case of strong Jahn-Teller coupling,
opposite to (\ref{relative}), one observes that if the
lattice distortion is standardly given by $Q_0=g_{JT}^2/\mu\omega_0^2$
then $g_{JT}Q_0$ is much larger then $T_K$.
So one deals with
a high field susceptibility when the pseudospin behaves as if
it is free. Hence the Kondo interaction (\ref{GeneralKondocoupling})
is irrelevant.
As it is well known\cite{Abraham}, the degeneracy of the
$Q_0^2=Q_\varepsilon^2+Q_\theta^2$ minima manifold  is lifted
by phonons anharmonicity and non-linear couplings (warping).
[Notice that the inequality (\ref{zero-point}) is still to be
satisfied.]
Resultingly, three equivalent minima are formed at the positions
$(0,1)$, $(\sqrt{3}/2,1/2)$, $(-\sqrt{3}/2,1/2)$ or at
the positions $(1/2,\sqrt{3}/2)$, $(0,-1)$, $(-1/2,\sqrt{3}/2)$
depending on the details of the warping potential.
With each of these minima a state of the system described by
a product of the phonon and the pseudospin wave functions
is associated.
The three degenerate states will be denoted
as $|1\rangle$,$|2\rangle$, and $|3\rangle$.
As a matter of fact, the overall cubic symmetry of
the total Hamiltonian (\ref{GeneralHamiltonian}) can not be
violated.
This is reflected by tunneling rates for tunneling of
the system between the states $|i\rangle$ ($i=1,2,3$) being finite.
In the conventional theory of the Jahn-Teller
effect\cite{Abraham,Bersuker,Ham}, a
diagonalization of the tunneling Hamiltonian leads to
``vibronic states": the ground doublet $\tilde{\Gamma}_3$, spanned
(in the lowest approximation in the tunneling rate) by the
wave functions $|\tilde{\varepsilon}\rangle=[2|1\rangle-
|2\rangle-|3\rangle]/\sqrt{6}$ and $|\tilde{\theta}\rangle=
[|2\rangle-|3\rangle]/\sqrt{2}$, and the excited singlet with the
wave function $|\tilde{\Gamma}_1\rangle=[|1\rangle+
|2\rangle+|3\rangle]/\sqrt{3}$ separated from the ground doublet
by an energy $\Delta E$
[of the order of $W_{JT}\exp(-3W_{JT}/2\omega_0)$].

It needs to be realized that in the case of an impurity ion
in a metal, the distorted configurations strongly interact with
conduction electrons.
This interaction has nothing to do with the
Kondo coupling (\ref{GeneralKondocoupling}) - it is caused by
the conventional electron-phonon interaction present in
(\ref{GeneralHamiltonian}).
Indeed, {\bf (i)} because of the latter interaction,
the conduction electrons experience different scattering potentials
when the impurity complex is in different distorted configurations.
Hence the screening effect.
One may expect the corresponding dimensionless coupling constant
to be of the order of unity (if the same is true for the bulk
electron-phonon coupling).
The role of the screening is to renormalize (decrease) the energy splitting
$\Delta E$.
Additionally, {\bf (ii)} the electron density fluctuations modify
tunneling barriers between the configurations $|i\rangle$.
This gives rise to assisted tunneling processes (analogously
to the case of two-level systems\cite{Zawadowski}).
These processes, though in several orders of magnitude smaller
then the screening interaction, lead to a nontrivial scaling
making the problem formally similar to the Kondo effect.
As a result, one is dealing with a ``Jahn-Teller-Kondo"
phenomenon when the local electronic configuration,
the lattice distortions and the conduction electrons form a
coherent state.
The coupling is of the form $\sum_{ij} \hat{V}_{ij} |i\rangle\langle j|$,
where $\hat{V}_{ij}$ are the electron scattering operators of
appropriate symmetry\cite{elsewhere}.
The conduction electrons can never be disregarded.
The present formulation gives rise to several low
energy scales\cite{notescales}.
At temperatures $\Delta E \ll T \ll W_{JT}$ the corresponding model
resembles a very anisotropic pseudospin-one model which critical
behaviour is to be studied\cite{foma}.
At the lowest temperatures ($T \ll\Delta E$) one again expects a
two channel behaviour (within the vibronic doublet) which can
however not be further quenched.

To conclude,
the problem of the interplay of the Kondo-
and the Jahn-Teller effects is formulated. Parameters regions where
the system exhibits qualitatively different properties
are determined. The $\Gamma_3$ ground doublet is discussed as an
example though the considerations
can also be applied, with minor adjustments\cite{noteadjustements},
to other
degenerate representations.
I believe that various ideas presented here
deserve further investigations.

I am thankful to Daniel Cox for helpful discussions and comments.


\begin{references}

\bibitem{single-Kondo}
K.G. Wilson, Rev. Mod. Phys. {\bf 47}, 773 (1975);
P.W. Anderson, G. Yuval and D.R. Hamann,
Phys. Rev. B {\bf 1}, 4664 (1970);
P. Nozi\`eres, J. Low Temp. Phys. {\bf 17}, 31 (1974);
N. Andrei, Phys. Rev. Lett. {\bf 45}, 379 (1980);
P.B. Wiegmann, JETP Lett. {\bf 31}, 364 (1980).

\bibitem{Anderson}
P.W. Anderson, Phys. Rev. {\bf 124}, 41 (1961).

\bibitem{Goqblin-Schrieffer}
B. Goqblin and J.R. Schrieffer, Phys. Rev. {\bf 185}, 847 (1969).

\bibitem{Hirst}
L.L. Hirst, Adv. Phys. {\bf 27}, 231 (1978).

\bibitem{Boss2}
P. Nozi\`eres and A. Blandin, J. Phys. (Paris) {\bf 41}, 193 (1980).

\bibitem{Bethe-multi}
N. Andrei and C. Destri, Phys. Rev. Lett. {\bf 52}, 364 (1984);
A.M. Tsvelik and P.B. Wiegmann, Z. Phys. B {\bf 54}, 201 (1984).

\bibitem{large-k}
J. Gan, N. Andrei and P. Coleman, Phys. Rev. Lett. {\bf 70}, 686 (1993);
D.L. Cox and A.E. Ruckenstein, Phys. Rev. Lett. {\bf 71}, 1613 (1993).

\bibitem{CFT}
A.M. Tsvelik, J. Phys. C {\bf 2}, 2833 (1990);
I. Affleck and A.W.W. Ludwig, Nucl. Phys. B {\bf 360}, 641 (1991).

\bibitem{bosonization}
V.J. Emery and S. Kivelson, Phys. Rev. B {\bf 47}, 10812 (1992);
D.G. Clarke, T. Giamarchi and B.I. Shraiman, {\em ibid.} {\bf 48},
7070 (1993);
A.M. Sengupta and A. Georges, {\em ibid.} {\bf 49}, 10020 (1994);
M. Fabrizio and A.O. Gogolin, Phys. Rev. B {\bf 50}, 17732 (1994);
M. Fabrizio, A.O. Gogolin, and P. Nozi\`eres, to appear in Phys. Rev. B.
For a discussion establishing a link between this approach and
Bethe-ansatz methods see A.M. Tsvelik, cond-mat/9502081.

\bibitem{Zawadowski}
A. Zawadowski, Phys. Rev. Lett. {\bf 45}, 211 (1980);
A. Muramatsu and F. Guinea, {\it ibid.} {\bf 57}, 2337 (1986);
G. Zar\'and and A. Zawadowski, {\it ibid.} {\bf 72}, 542 (1994).

\bibitem{PCS}
D.C. Ralph and R.A. Buhrman, Phys. Rev. Lett. {\bf 69}, 2118 (1992);
D.C. Ralph, A.W.W. Ludwig, Jan von Delft and R.A. Buhrman,
{\it ibid.} {\bf 72}, 1064 (1994).

\bibitem{hf-general}
See, e.g., F. Steglich, C. Geibel, K. Gloos, G. Olesch, C. Schrank,
C. Wassilew, A. Loidl, A. Krimmel, and G.R. Stewart,
J. Low Temp. Phys. {\bf 95}, 3 (1994).

\bibitem{Cox1}
D.L. Cox, Phys. Rev. Lett. {\bf 59}, 1240 (1987).
For experimental data see
C.L. Seaman, M.B. Maple, B.W.Lee, S. Ghamaty, M.S. Torikachvili,
J.-S. Kang, L.Z. Lui, J.W. Allen, and D.L. Cox, {\it ibid.} {\bf 67},
2882 (1991), but also B. Andraka and A.M. Tsvelik, {\it ibid.}
{\bf 67}, 2886 (1991).

\bibitem{Cox2}
B. Andraka, Phys. Rev. B {\bf 49}, 3589 (1994);
Tae-Suk Kim and D.L. Cox, cond-mat/9412024.

\bibitem{Maple}
M.B. Maple, C.L. Seaman, D.A. Gajewski, Y. Dalichaouch, V.B. Barreta,
M.C. de Andrade, H.A. Mook, H.G. Lukefahr, O.O. Bernal,
and D.E. MacLaughlin,
J. Low. Temp. Phys. {\bf 95}, 225 (1994).

\bibitem{JimAllen}
J.W. Allen, L.Z. Liu, R.O. Anderson, C.L. Seaman, M.B. Maple,
Y. Dalichaouch, J.-S. Kang, M.S. Torikachvili and
M.L. de la Torre, Phys. B {\bf 186-188}, 309 (1993).

\bibitem{splittings}
K.R. Lea, M.J.M. Leask, and W.P. Wolf, J. Phys. Chem. Solids
{\bf 23}, 1381 (1962).

\bibitem{neutron}
H.A. Mook, C.L. Seaman, M.B. Maple, M.A. Lopez de la Torre, D.L. Cox,
and M.Makivic, Phys. B {\bf 186-188}, 341 (1993).

\bibitem{Cox3}
D.L. Cox and M. Makivic,
Phys. B {\bf 199-200}, 391 (1994).

\bibitem{Abraham}
A. Abraham and B. Bleaney,
{\em Electron paramagnetic resonance of transition ions},
Clarendon Press, Oxford, 1970.

\bibitem{elsewhere}
A.O. Gogolin, unpublished.

\bibitem{Cox4}
D.L. Cox, Phys. C {\bf 153}, 1642 (1988).

\bibitem{Sacramento}
D. Sacramento and P. Schlottmann, Phys. Lett. A {\bf 142}, 245 (1989).

\bibitem{noteperturbation}
Yet it will introduce $g$-factors, modifying the coupling
of the $\Gamma_3$ doublet with a strain field, etc.

\bibitem{YuAnderson}
C.C. Yu and P.W. Anderson, Phys. Rev. B {\bf 29}, 6165 (1984).

\bibitem{Bersuker}
I.B. Bersuker, Z. Eksp. Teor. Fiz. {\bf 43}, 1315 (1962)
[Sov. Phys. JETP {\bf 16}, 933 (1963)];
Z. Eksp. Teor. Fiz. {\bf 44}, 1239 (1963)
[Sov. Phys. JETP {\bf 17}, 836 (1963)].

\bibitem{Ham}
F.S. Ham, Phys. Rev. {\bf 166}, 307 (1968).

\bibitem{notescales}
So, for $U-MPd_3$ systems, the experimental distinction of the $M=La$
case from $M=(Y,Sc)$ cases
(see \protect\cite{Maple}\protect)
may be related to the larger Fermi level
density of states for the $La$ compounds [as follows from
band structure calculations by C. K\"onig, Z. Phys. B
{\bf 50}, 33 (1983)] that in turn effects the balance of the Kondo-
and the Jahn-Teller couplings.

\bibitem{foma}
Entirely different but also involving lattice degrees of freedom
impurity models have been discussed by H. Capellmann and
A.S. Ioselevich (to appear in Phys. Rev. B).

\bibitem{noteadjustements}
These will mainly concern the nature and the low energy behaviour of
the pseudospin magnetization.

\end{references}
\end{document}